# Thin film growth of a topological crystal insulator SnTe on the CdTe (111) surface by molecular beam epitaxy

Ryo Ishikawa[1], Tomonari Yamaguchi[1], Yusuke Ohtaki[1], Ryota Akiyama[2], and Shinji Kuroda*[1]

[1] Institute of Materials Science, University of Tsukuba, 1-1-1 Tennoudai, Tsukuba, Ibaraki 305-8573, Japan

[2] Department of Physics, Faculty of Science, The University of Tokyo, 7-3-1 Hongo, Bunkyo-ku, Tokyo 113-0033, Japan

## Abstract

We report molecular beam epitaxial growth of a SnTe (111) layer on a CdTe template, fabricated by depositing it on a GaAs (111)A substrate, instead of $BaF_2$ which has been conventionally used as a substrate. By optimizing temperatures for the growth of both SnTe and CdTe layers, we could obtain SnTe layers of the single phase grown only in the (111) orientation and of much improved surface morphology from the viewpoint of the extension and the flatness of flat regions, compared to the layers grown on $BaF_2$. In this optimal growth condition, we have also achieved a low hole density of the order of $10^{17}cm^{-3}$ at 4K, the lowest value ever reported for SnTe thin films without additional doping. In the magnetoresistance measurement on this optimized SnTe layer, we observe characteristic negative magneto-conductance which is attributed to the weak antilocalization effect of the two-dimensional transport in the topological surface state.

## 1. Introduction

Topological crystalline insulator (TCI) has been attracting attention as a novel class of topological insulators (TIs), in which the topological surface state (TSS) is protected by the mirror-reflection symmetry of the crystal[1,2], instead of the time-reversal symmetry in the case of conventional TIs. SnTe and SnSe are known as typical materials of TCIs, with the topological protection is due to the mirror reflection symmetry with respect to the set of six {110} planes of the rock-salt crystal structure. Experimentally the topological nature of SnTe(Se) and related mixed crystals was confirmed by the observation of the Dirac-shape surface band in the angle-resolved photoemission spectroscope (ARPES) measurement[3,4]. Electrical transport properties characteristic of the TSS have also been demonstrated in SnTe nanowires[5] or thin films[6-9]. But a large contribution of the bulk transport, originating from heavily p-doped tendency due to the formation of high-density Sn vacancies[10], sometimes hinders us from extracting the properties of the 2D transport in the TSS. Thin films have the advantage in reducing a bulk contribution due to an enhanced surface-to-bulk ratio, but another problem of the surface morphology arises in the case of thin films deposited on a different material for substrate. For the growth of SnTe thin films, $BaF_2$ has conventionally been used as a substrate material due to a small lattice mismatch ratio(lattice mismatch ratio 1.7 %)[11,12]. However, it has been reported that SnTe thin films grown on $BaF_2$ exhibit a rough surface of a textured structure, with pronounced island-like features particularly for a small film thickness below 100 nm[7,8]. This is because the growth of SnTe proceeds by the nucleation and island growth, with the formation of boundaries between islands, which may hinder the carrier transport on the TSS.

In the present study, we choose CdTe as an alternative material of substrate for the growth of SnTe. The growth on CdTe gives rise to a slightly larger lattice mismatch with SnTe (mismatch ratio 2.8%) compared to the growth on $BaF_2$. However, it may be advantageous for the growth of SnTe with a high quality that SnTe and CdTe belong the same group of fcc crystal structure and have the common anion element. In the present study, we have grown SnTe (111) layers on CdTe by molecular beam epitaxy (MBE) with a systematic variation of growth conditions and have investigated how morphology, crystallinity, and transport properties change depending on the growth conditions.

## 2. Experimental methods

In this study, we used a CdTe(111) template, a thick CdTe layer deposited on a GaAs(111)A wafer, as a substrate for the growth of a SnTe(111) layer. First, we fabricated a CdTe template by depositing CdTe on a GaAs wafer in an MBE chamber equipped with elemental sources of Cd, Zn, and Te. After cleaning the surface of a GaAs wafer using rf-excited hydrogen plasma, we first deposited a thin ZnTe layer with a few-nm thickness and then deposited a thick CdTe layer with a thickness of 500-600 nm. The temperature of the substrate was kept at a fixed value of 315$^{o}$C during the growth of ZnTe, and changed in the range of 180 – 240$^{o}$C during the growth of CdTe. Both of the ZnTe and CdTe layers were grown with a surplus supply of Te flux over Zn or Cd flux (Te-rich flux condition). The growth of SnTe on a CdTe template was performed in another MBE chamber equipped with a compound source of SnTe. In the growth of SnTe, the amount of SnTe flux was fixed at a constant value and the temperature of the substrate was changed in the range of 240 – 300$^{o}$C. For reference, we also grew SnTe on a $BaF_2$ (111) substrate. The surface was monitored by reflection high energy electron diffraction (RHEED) during the growth of the respective layers. After the growth, the surface morphology was examined by atomic force microscope (AFM) using the dynamic mode in the air. The crystal structure and orientation was checked by the $\theta$-2$\theta$ scan of x-ray diffraction (XRD). We performed the electrical transport measurement on the grown films using Physical Properties Measurement System (PPMS). The measurement was done in the standard six-probe configuration with electrodes fabricated by soldering indium on the surface of the SnTe layer.

## 3. Experimental results and discussion

First, we checked the surface of the substrate before the growth of SnTe. Figure 1 shows RHEED images of the surface of the two substrates, the $BaF_2$ surface just after annealed at 300$^{o}$C and the surface of CdTe grown at $T_{CdTe}$ = 240$^{o}$C. One can see a clear contrast between the two RHEED images; a spotty pattern of the $BaF_2$ and a streaky pattern of the CdTe, indicating a rough and a flat surface, respectively. Figure 1 (c) shows the AFM image of the same CdTe surface shown in Fig. 1 (b), which exhibits an almost flat surface with the root mean square (rms) of height of 1.5 nm. Next, we grew SnTe on a CdTe template and investigated how the surface morphology of the SnTe layer changes depending on the growth temperature of the CdTe template $T_{CdTe}$ and that of the SnTe layer $T_{SnTe}$. Figure 2 compares AFM images of the surface of SnTe layers grown on CdTe templates at different growth temperatures of $T_{CdTe}$ and $T_{SnTe}$, together with the surface of SnTe grown on a $BaF_2$ substrate; in Fig. 2 (b)-(d), $T_{CdTe}$ is varied in the range of $T_{CdTe}$ = 195 – 240°C at a fixed value of $T_{SnTe}$ = 240°C, and in Fig. 2 (d)-(f), $T_{SnTe}$ is varied in the range of $T_{SnTe}$ = 240 – 300°C at a fixed value of $T_{CdTe}$ = 240°C. In all these samples, the thickness of the SnTe layers is around 50 nm. The surface of SnTe grown on $BaF_2$ exhibits a maze-like textured structure, consisting of winding stripes bordered with deep ditches. This structure is considered to be formed by the coalescence of each island into continuous stripes in the nucleation and island growth mode. In the SnTe surface grown on $BaF_2$ shown in Fig. 2 (a), the characteristic sizes of this textured structure are given in the following; the breadth of each stripe is around 100 nm, the rms of height within each stripe around 6 nm, and the depth of ditches is 10 – 30 nm. The SnTe surface grown on CdTe exhibits a similar textured structure, but the breadth of each stripe becomes larger. When we compare the SnTe surface on CdTe grown at different $T_{CdTe}$, shown in Fig. 2 (b)-(d), the breadth of each stripe increases with the increase of $T_{CdTe}$ from ~ 200 nm at $T_{CdTe}$ = 195°C to ~300 nm at $T_{CdTe}$ = 240°C. On the other hand, the rms of height within each stripe is around 20 nm at a low growth temperature of $T_{CdTe}$ = 195°C, but with the increase of $T_{CdTe}$, the rms decreases down to ~2 nm at $T_{CdTe}$ = 240°C. The depth of ditches does not change much, remaining at around 30 nm. These differences of the SnTe surface reflect the surface morphology of the CdTe template grown at different temperature; the RHEED image of the CdTe surface exhibits more streaky feature in the growth at a higher value of $T_{CdTe}$ (not shown), indicating the surface becomes more flat. When we compare the SnTe surface grown at different $T_{SnTe}$, shown in Fig. 2 (d)-(e), the breadth of stripes becomes larger with increasing $T_{SnTe}$ from 240 to 270°C, but the rms of height within each stripe also increases from ~ 2nm at $T_{SnTe}$ = 240°C to ~ 10 nm at $T_{SnTe}$ = 270°C. When $T_{SnTe}$ increases up to 300°C, it looks like the growth mode changes, the surface consisting of terraces and holes of triangular shape, instead of stripes and ditches, but the rms of height within each terrace decreases down to ~ 4 nm.

The crystal structure of the grown SnTe layers was examined by XRD. Figure 3 shows $\theta$-$2\theta$ scan profiles of SnTe layers grown on CdTe templates at different values of $T_{CdTe}$ and $T_{SnTe}$. As shown in the figure, for the SnTe layers grown at $T_{SnTe}$ = 240°C, any other diffraction peaks were not detected than those of the SnTe (111) planes, irrespective of the values of $T_{\mathrm{CdTe}}$, in addition to the diffractions from GaAs and CdTe. On the other hand, when $T_{\mathrm{SnTe}}$ is increased to 270 and 300°C, additional diffraction peaks start to appear, as denoted by black hollow arrows in the figure. These peaks are assigned to the diffractions from the SnTe (100) plane. This result indicates that the SnTe layer grown on CdTe at $T_{\mathrm{SnTe}}$ = 240°C is oriented only in the (111) crystallographic orientation, but at higher temperatures of $T_{\mathrm{SnTe}}$ = 270 – 300°C, the SnTe layer consists of the mixture with regions grown in the (100) orientation. As described above, the characterizations of morphology and crystallinity using AFM and XRD allow us to conclude that the optimal temperature for the growth of both CdTe and SnTe is $T_{\mathrm{CdTe}} = T_{\mathrm{SnTe}}$ = 240°C from the viewpoint of the surface morphology, in particular, the degree of extension of flat regions, the flatness within the respective flat regions, and crystallographically single phase.

We performed electric transport measurements on the SnTe films characterized above. Figure 4 shows the temperature dependence of resistivity of the SnTe layers on CdTe templates grown respectively at $T_{\mathrm{CdTe}}$ = 240°C and $T_{\mathrm{SnTe}}$ = 240 – 300°C. All these films exhibit metallic behaviors with a monotonic decrease of resistivity with lowering temperature, but the value of resistivity is much different by the growth temperature $T_{\mathrm{SnTe}}$. This difference in resistivity is attributed to the difference in carrier concentration, as demonstrated by the Hall measurement described below. We measured the electric transport under magnetic fields applied perpendicular to the film plane. From the longitudinal (sheet) resistance $R_{xx}$ and the Hall resistance $R_{xy}$ under magnetic fields, we deduce the Hall conductance $G_{xy}$ using equation $G_{xy} = R_{xy}/(R_{xx}^2 + R_{xy}^2)$. Figure 5 shows thus deduced $G_{xy}$ as a function of the magnetic field $B$ at 4 K. As shown in the figure, $G_{xy}$ exhibits non-linear dependence on $B$. The deviation from the expected linear dependence is attributed to a large value of carrier mobility. To fit these non-linear curves of $G_{xy}(B)$, we used the Drude expression of the Hall conductivity

$$G_{xy} = \frac{n_{\mathrm{H}} e \mu_{\mathrm{H}}^2 B}{1 + (\mu_{\mathrm{H}} B)^2}$$

where $e$ is the charge of an electron. The Hall carrier density $n_{\mathrm{H}}$ and mobility $\mu_{\mathrm{H}}$ obtained from the fitting are listed in Table 1. The hole density $n_{\mathrm{H}} = 2.05 \times 10^{17}\mathrm{cm}^{-3}$ in the film grown at $T_{\mathrm{SnTe}}$ = 240°C is much smaller compared to the smallest value of $n_{\mathrm{H}} = 1.0 \times 10^{18}\mathrm{cm}^{-3}$ reported for SnTe films grown on a $BaF_2$ substrate[9]. With the increase of $T_{SnTe}$, the value of $n_H$ increases significantly while the mobility does not change much. A significant increase of $n_H$ with $T_{SnTe}$ could be attributed to the formation of Sn vacancies, which act as acceptors in SnTe and supply holes. The density of Sn vacancies is much enhanced in the growth at a higher temperature.

We have also analyzed the magnetoresistance behavior in order to extract features of the surface transport. We derive the magneto-conductance (MC) by subtracting the conductance at zero field from the value at a given field $B$ as $\Delta G(B) = G(B) - G(0) = 1/R_{xx}(B) - 1/R_{xx}(0)$. Figure 6 shows the plot of $\Delta G(B)$ as a function of $B$ in the both configurations of magnetic fields parallel and perpendicular to the film plane. For the film grown at $T_{SnTe}$ = 240°C, a cusp-like structure appears around zero field in the MC curves in the perpendicular field configuration (perpendicular MC) $\Delta G_{\perp}(B)$, that is, a rapid decrease in conductance with the application of a small magnetic field. This negative MC is attributed to the WAL effect of the 2D transport on the TSS[13,14], a quantum correction to conductivity in the diffusive transport regime arising from the interference of wavefunctions[15]. The π shift in the Berry phase in the TSS transport causes a destructive interference of wavefunctions along a closed trajectory, leading to an enhancement of conductivity[16]. This interference effect falls off upon the application of a magnetic field perpendicular to the trajectory, leading to a reduction of conductivity with magnetic field. Since the reduction of conductivity should be effective only when a magnetic field is applied perpendicular to the transport surface, the negative MC component due to the WAL effect in the TSS can be extracted by subtracting the MC under a parallel field $\Delta G_{//}(B)$ (parallel MC) from the perpendicular MC $\Delta G_{\perp}(B)$, as is often done in the analysis of the WAL effect[9,14]. In the lower panel of Fig. 6, the subtracted MC, $\Delta G_{\perp}(B) - \Delta G_{//}(B)$ is plotted against $B$. In the film grown at $T_{SnTe}$ = 240°C, the subtracted MC curve consists of a sharp peak around zero field and a broad upward convex curve. The sharp peak is attributed to the reduction of MC due to the WAL effect in the TSS. We fit this sharp peak using Hikami-Nagaoka-Larkin (HNL) equation[15], which gives the MC originating from the quantum correction of the 2D diffusive transport, as follows.

$$\Delta\sigma = \sigma(B) - \sigma(0) = -\frac{\alpha e^2}{2\pi^2\hbar}\left[\ln\left(\frac{\hbar}{4Bel_\phi{}^2}\right) - \Psi\left(\frac{1}{2} + \frac{\hbar}{4Bel_\phi{}^2}\right)\right]$$

Here, $\hbar$ is Planck's constant, $l_\phi$ is the phase coherence length, $\Psi(x)$ is the digamma function, and $\alpha$

represents a coefficient related to the number of coherent transport channels. As a result of fitting the experimental curve to this equation, we obtain the values of the fitting parameters $l_\phi$ and $\alpha$ as $l_\phi$ = 322 nm and $\alpha$ = -0.025. Since it is known that the value of $\alpha$ = -0.5 should be allotted per single transport channel in the TSS, the number of transport channel is given by 2|$\alpha$| = 0.05. This extremely small number may be due to the competition with a positive MC component arising from the weak localization (WL) effect in the bulk subbands, which were formed by the confinement along the thickness of the film; the negative MC component due to the WAL effect in the TSS is reduced significantly by a superimposed positive MC component of the WL effect in the bulk subbands[17]. In the films grown at higher temperatures of $T_{SnTe}$ = 270, 300°C, the perpendicular MC curves $\Delta G_\perp(B)$ exhibit a pronounced cusp around zero field, but a similar cusp-like structure is also seen in the parallel MC curve $\Delta G_{//}(B)$ though the width of the cusp is broadened compared to that in $\Delta G_\perp(B)$. From a similarity of the cusp shape, it would be reasonable to assume that the negative MC observed both in the perpendicular and parallel MC curves should have the same origin. It could not be attributed to the 2D WAL effect, for which the suppression of conductivity should be effective only for the application of magnetic field perpendicular to the surface plane. It would rather be reasonable to attribute it to the 3D WAL effect in the bulk transport. Even in the bulk conduction, the WAL effect should be expected to arise in the existence of a large spin-orbit coupling, then the application of magnetic field results in the suppression of conductivity irrespective of the direction of a magnetic field. In the transport in a thin film, 2D subbands are formed from the bulk band due to the confinement along the thickness of the film. These 2D subbands are expected to give rise to anisotropic behaviors in the positive MC since the suppression of the WAL effect should be different between the perpendicular and parallel field configurations due to anisotropic characters of the subbands[18]. As demonstrated in $Bi_2Se_3/In_2Se_3$ superlattices[18], the cusp in the parallel MC becomes wider than that in the perpendicular MC due to the anisotropy of these 2D subbands. In the films grown at $T_{SnTe}$ = 270, 300°C, having high hole densities, a number of 2D subbands are populated, which contribute to the conduction as independent transport channels, resulting in an enhanced 3D WAL effect. This 3D WAL effect becomes dominant in the observed MC curves, giving the anisotropic negative MC with a large magnitude. On the other hand, in the film grown at $T_{SnTe}$ = 240°C with a low hole density, the 3D WAL is considered to be ineffective since these 2D subbands are not well populated. In addition, the crossover from WAL to WL effect might be expected with the decrease of the Fermi energy relative to the band-gap[17]. This would give a reasonable explanation for the observed positive small MC peak due to the 2D WAL effect of the surface transport only in the film grown with a low hole density.

## 4. Summary

We have performed the growth of a SnTe (111) layer on CdTe template grown on a GaAs (111)A substrate. The surface morphology and crystallinity were characterized using AFM and XRD on the series of the films grown at different growth temperatures of CdTe and SnTe. Though the surface of the SnTe layer grown on a CdTe exhibits a textured structure, similarly to SnTe grown on $BaF_2$, we achieve a much improvement of the surface flatness and the single phase grown only in the (111) orientation at an optimal temperature of 240°C both for the growth of CdTe and SnTe. We also achieve a low hole density of the order of $10^{17}$ $cm^{-3}$ at 4 K at this optimal growth temperature. In the magnetoresistance measurement, we observe a negative magneto-conductance due to the WAL effect, which is characteristic of the 2D transport in the TSS.

## Acknowledgements

The authors would like to thank T. Koyano in the Cryogenic Division of Research Facility Center for Science and Technology, University of Tsukuba for PPMS measurements and T. Suemasu for AFM and XRD measurements. This work has partially been supported by Grand-in-Aid for Scientific Research (KAKENHI) from the Japan Society for the Promotion of Science.

*Corresponding author. E-mail address: kuroda@ims.tsukuba.ac.jp

**Figure captions**

Figure 1. (a), (b) RHEED images of the surface of two types of substrates before the growth of SnTe, (a) a $BaF_2$ substrate and (b) a CdTe template grown at $T_{CdTe}$ = 240°C, respectively. (c) AFM image of the same CdTe surface shown in (b).

Figure 2. AFM images of the surface of SnTe layers grown on CdTe templates at different growth temperatures of $T_{CdTe}$ and $T_{SnTe}$ (b)-(f), together with the surface of SnTe grown on a $BaF_2$ substrate (a). (b)-(d) : $T_{CdTe}$ is varied in the range of $T_{CdTe}$ = 195 – 240°C at a fixed value of $T_{SnTe}$ = 240°C. (d)-(f) : $T_{SnTe}$ is varied in the range of $T_{SnTe}$ = 240 – 300°C at a fixed value of $T_{CdTe}$ = 240°C.

Figure 3. XRD $\theta$-2$\theta$ scan profiles of SnTe layers grown on CdTe templates at different values of $T_{CdTe}$ and $T_{SnTe}$. The first three curves from the top : $T_{CdTe}$ is varied in the range of $T_{CdTe}$ = 195 – 240°C at a fixed value of $T_{SnTe}$ = 240°C. The remaining two curves: $T_{SnTe}$ is increased to 270, 300°C at a fixed value of $T_{CdTe}$ = 240°C. Small peaks denoted by black hollow arrows are assigned to the diffractions from SnTe (100) plane.

Figure 4. Plots of resistivity as a function of temperature for the SnTe layers grown on CdTe templates at $T_{CdTe}$ = 240°C and $T_{SnTe}$ = 240 – 300°C. The inset shows the same plot in an enlarged scale of the vertical axis.

Figure 5. Hall conductance $G_{xy}$ as a function of the applied magnetic field at 4 K for the same SnTe samples shown in Fig. 4.

Figure 6. Magneto-conductance (MC) under perpendicular magnetic fields $\Delta G_{\perp}(B)$ (red) and that under parallel magnetic fields $\Delta G_{//}(B)$ (black) (upper), and the subtracted MC $\Delta G_{\perp}(B) - \Delta G_{//}(B)$ (blue, lower) for the same SnTe samples shown in Fig. 4. The measurement temperature is 4 K.

**Table 1**

Hall carrier density ($n_H$) and mobility ($\mu_H$) at 4 K of the SnTe layers grown on CdTe templates at $T_{CdTe}$ = 240°C and $T_{SnTe}$ = 240 – 300°C. These values are obtained from the fitting of the curves of the Hall conductance $G_{xy}$ as a function of the magnetic field $B$ (Fig. 5).

| $T_{CdTe}$ (°C) | $T_{SnTe}$ (°C) | Carrier density $n_H$ ($cm^{-3}$) | Mobility $\mu_H$ ($cm^2$/V·sec) |
|---|---|---|---|
| 240 | 240 | $2.05 \times 10^{17}$ | 831 |
| | 270 | $2.46 \times 10^{18}$ | 804 |
| | 300 | $3.69 \times 10^{20}$ | 992 |

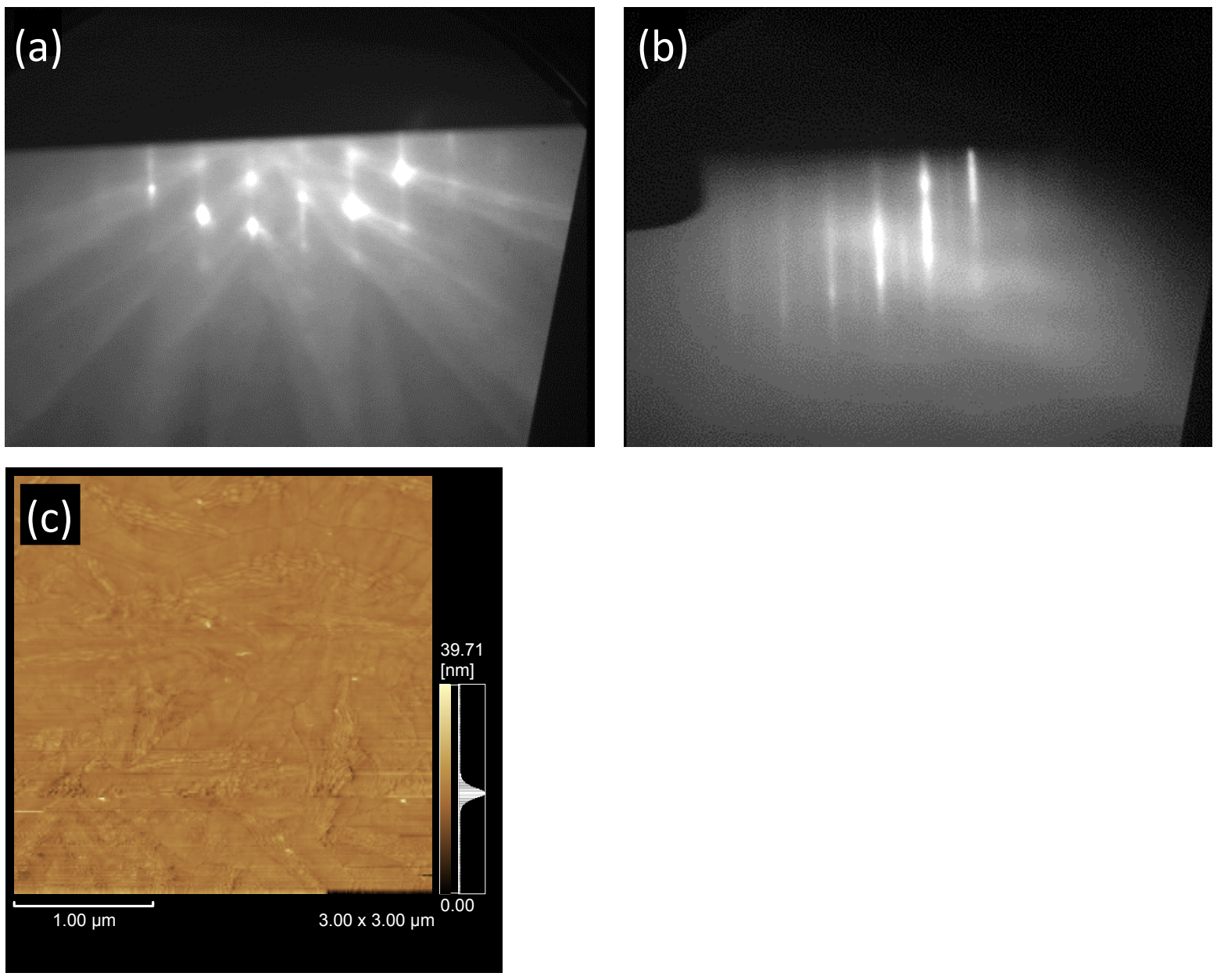

**Figure 1**

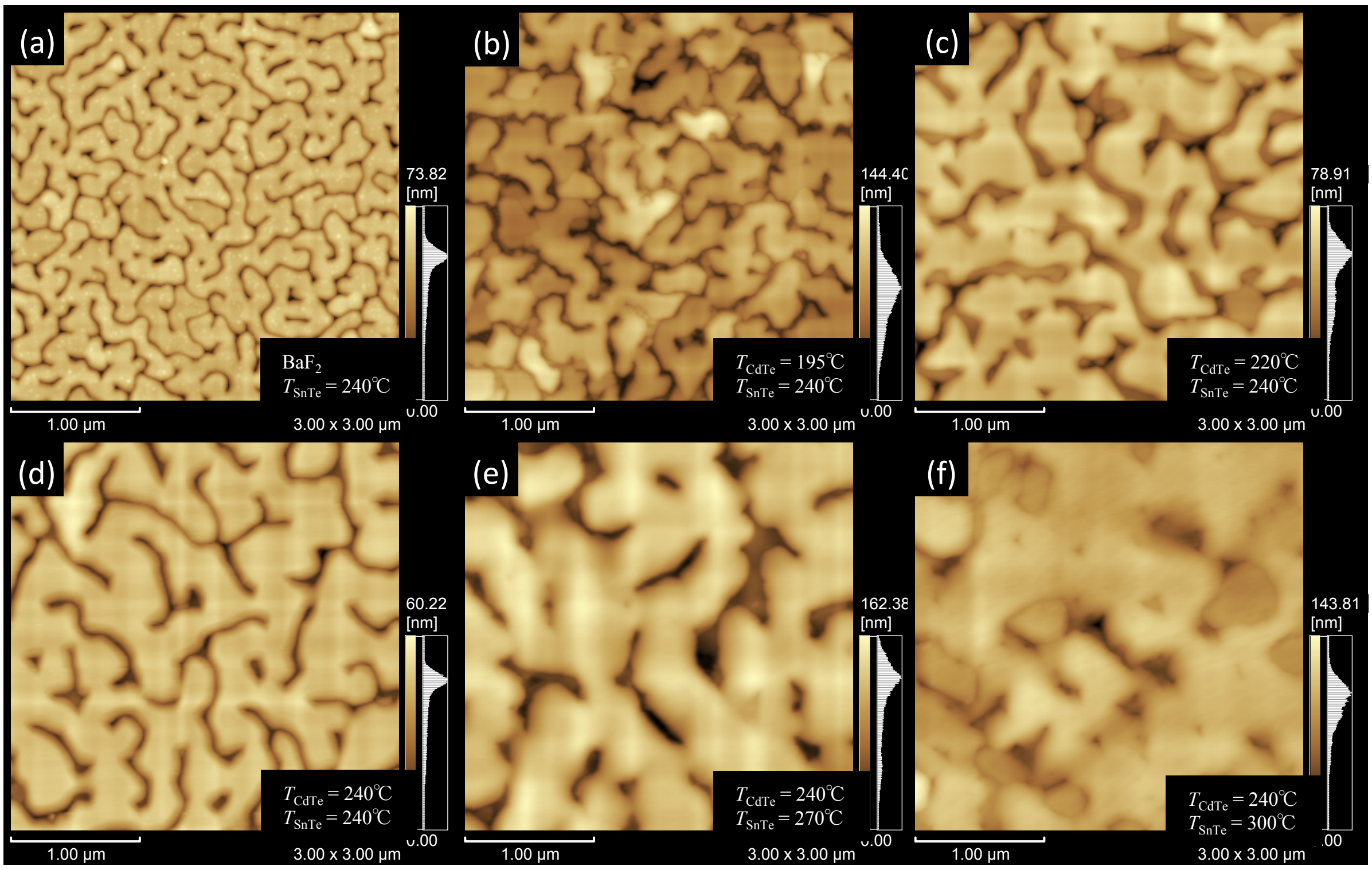

**Figure 2**

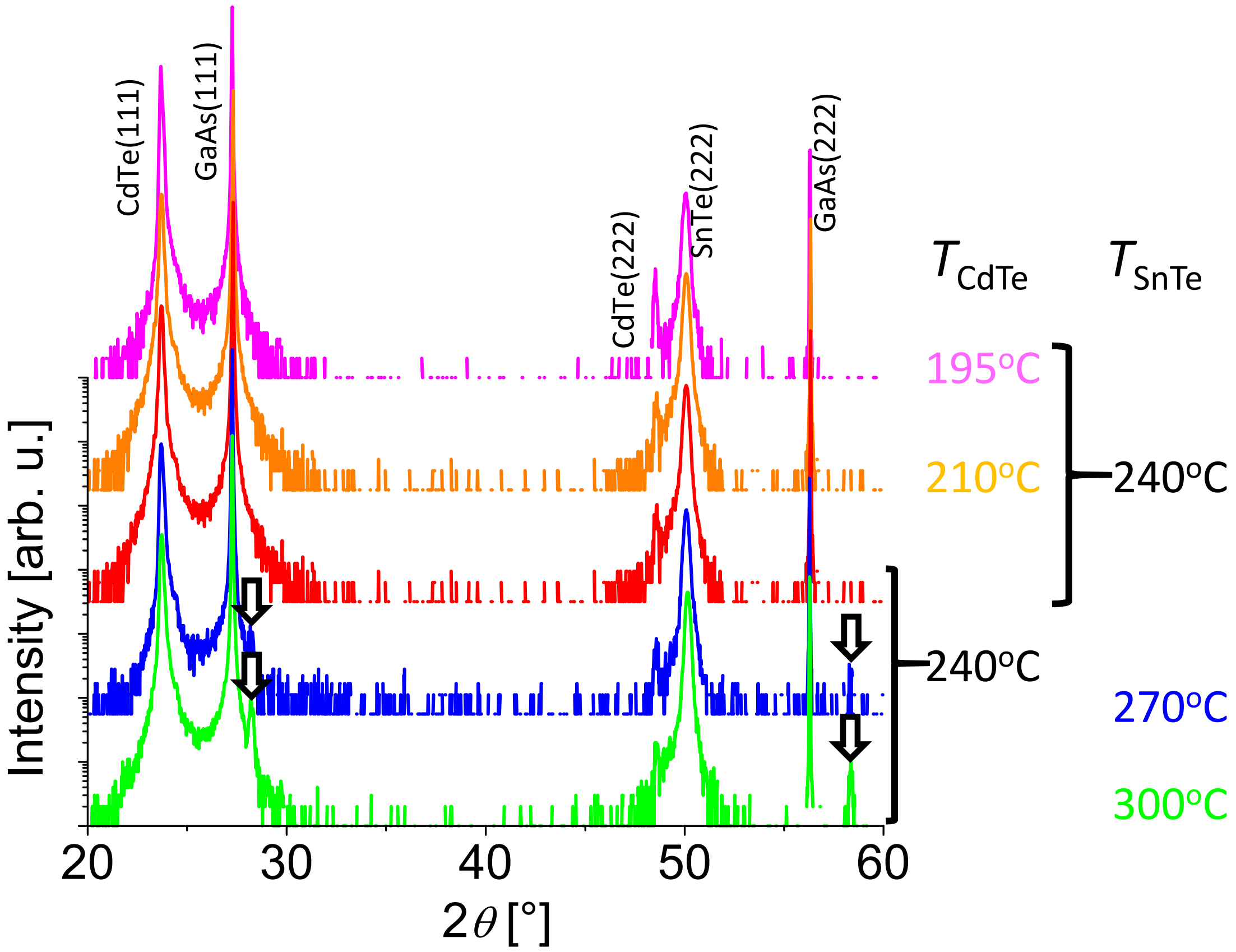

**Figure 3**

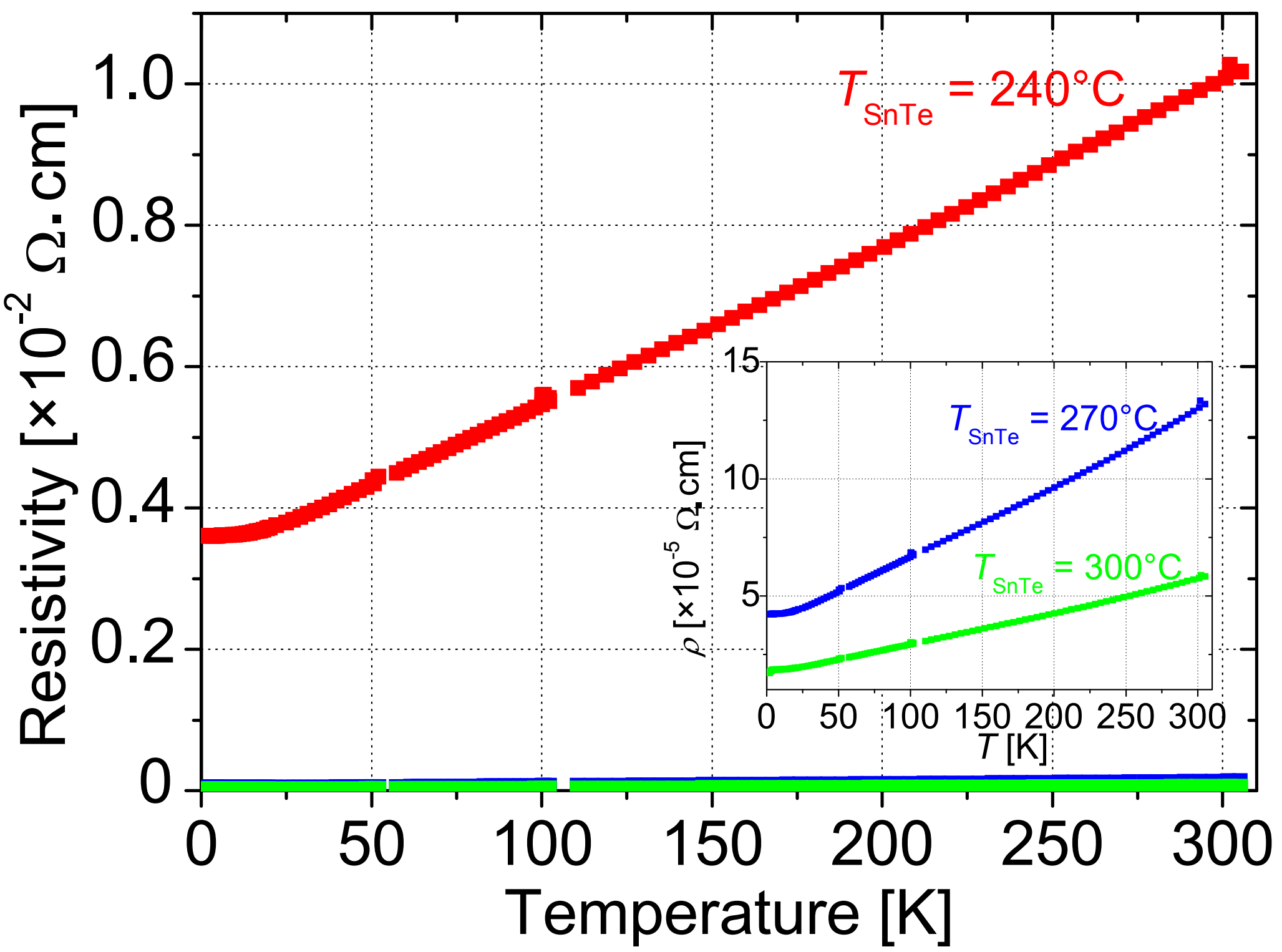

**Figure 4**

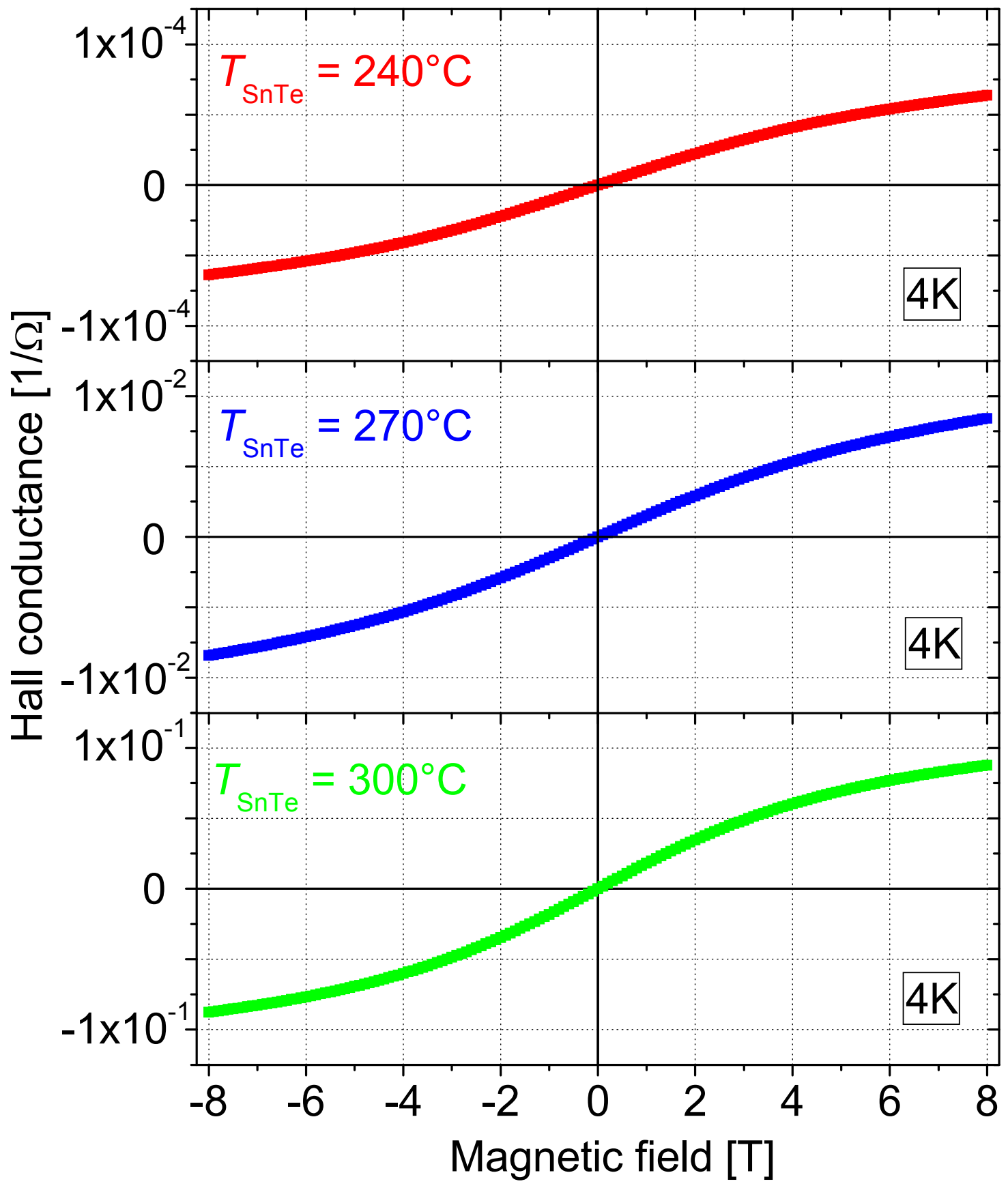

**Figure 5**

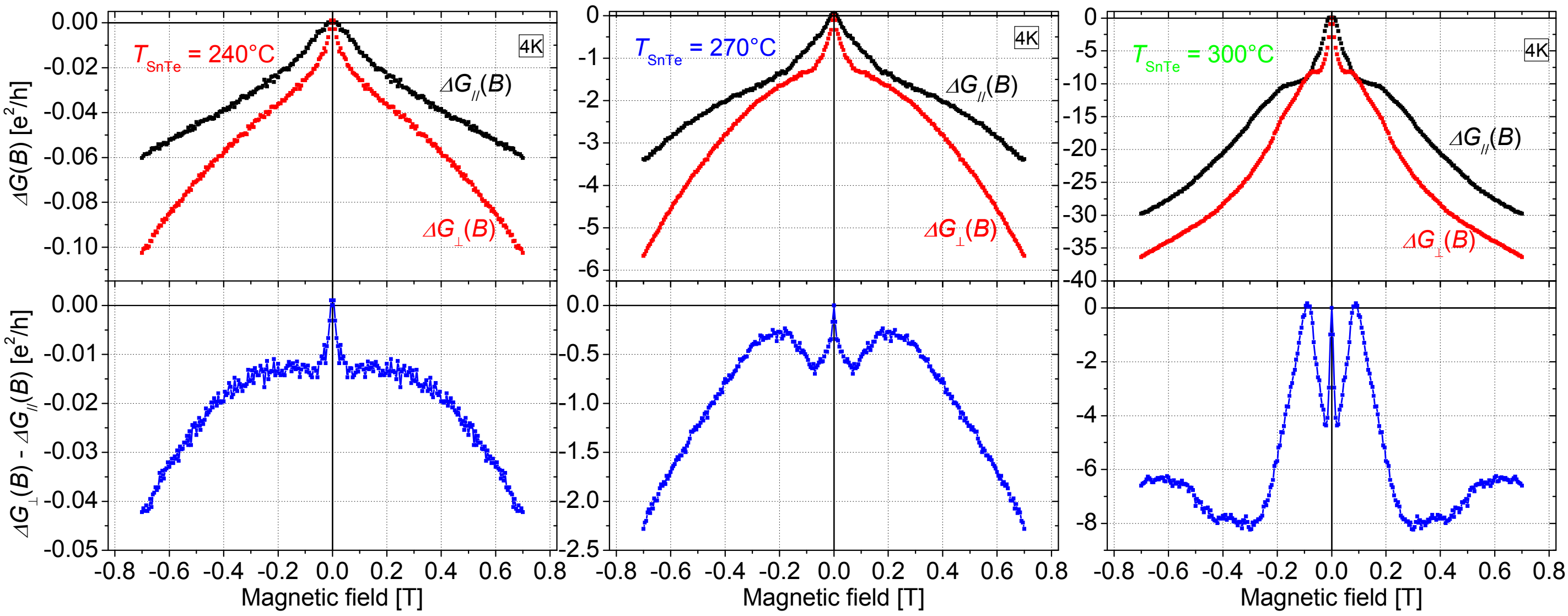

**Figure 6**